\documentclass[a4paper,12pt]{article}

\usepackage[T1]{fontenc}
\usepackage{lmodern}
\usepackage{textcomp}
\usepackage[utf8]{inputenc}
\usepackage{amsmath}
\usepackage{amssymb}
\usepackage{microtype}

\usepackage{hyperref}

\usepackage{natbib}
\bibpunct[:\ ]{(}{)}{;}{a}{}{,}

\title{The Hole Argument\thanks{Written for E.~Knox and A.~Wilson (eds), \emph{The Routledge Companion to the Philosophy of Physics} (Routledge, forthcoming). Please cite the published version.}%
}
\author{Oliver Pooley\thanks{email: oliver.pooley@philosophy.ox.ac.uk}%
\\
Oriel College, Oxford}
\date{21 September, 2020}

\begin{document}

\maketitle

\section{Introduction}

Our best theory of space, time and gravity is the general theory of relativity (GR). It accounts for gravitational phenomena in terms of the curvature of spacetime. In more mathematical presentations of the theory, solutions are standardly represented as n-tuples: $(M,g_{ab},\phi_1,\phi_2,\ldots)$. The \( \phi \)s are objects that represent the assorted material content of spacetime (such as stars and electromagnetic fields). $M$ and $g_{ab}$ together represent spacetime itself. $M$ is a differentiable manifold representing the 4-dimensional continuum of spacetime points. $g_{ab}$ is a Lorentzian metric tensor defined on $M$. It encodes some of spacetime's key spatiotemporal properties, such as the spacetime distances along paths in $M$. In particular, spacetime's curvature can be defined in terms of $g_{ab}$.

In its contemporary guise, the hole argument targets a natural interpretation of this mathematical machinery. According to the \emph{spacetime substantivalist}, spacetime itself, represented by $(M,g_{ab})$, should be taken to be an element of reality in its own right, on (at least) equal ontological footing with its material content. John Earman and John Norton's version of the hole argument \citep{earmannorton87} aims to undermine this reading of the theory. In particular, it seeks to establish that, under a substantivalist interpretation, GR is radically---and problematically---\emph{indeterministic}.

Earman and Norton’s hole argument set the agenda for the wide-ranging debate that burgeoned in the late 1980s and early '90s, and that has rumbled on in the decades since. The hole argument, however, did not originate with them. It was first advanced by Einstein \citep{einstein14com}, as he struggled to find a theory of gravity compatible with the notions of space and time ushered in by his 1905 special theory of relativity.

By 1913 Einstein had already settled on a non-flat metric tensor, $g_{ab}$, as the mathematical object that would capture gravitational effects. The remaining task was to discover field equations describing how $g_{ab}$ depends on material ``sources'', encoded by their stress energy tensor $T_{ab}$
.

Einstein sought a theory that would generalise the \emph{relativity principle}. The restricted relativity principle of Newtonian physics and special relativity only asserts the equivalence of all \emph{inertial} frames: frames in which (\emph{inter alia}) force-free bodies move uniformly in straight lines. The inertial frames are, however, physically distinguished in these theories from frames moving non-uniformly with respect to them. Einstein believed that fundamental physics should treat \emph{all} frames of reference as on a par.

In 1913, Einstein came tantalisingly close to settling on the famous field equations that he would eventually publish towards the end of 1915. These equations are \emph{generally covariant}: they hold in all coordinates systems within a family related by smooth but otherwise arbitrary coordinate transformations. Because such transformations include transformations between coordinates adapted to frames in arbitrary states of motion, Einstein initially believed 
that generally covariant equations embody a generalised relativity principle.

In 1913, however, he temporarily gave up the quest for general covariance. His original version of the hole argument convinced him that any generally covariant theory describing how $g_{ab}$ relates to $T_{ab}$ must be indeterministic. (For further details of Einstein's argument and its role in his search for his field equations, see \cite{stachel89}, \citet{norton84how} and \citet[\S3]{janssencc}.)

In this chapter, I focus on the hole argument as an argument against substantivalism. The next section reviews some technical notions standardly presupposed in presentations of the argument. Section~\ref{s:argument} presents the argument itself. The remainder of the chapter reviews possible responses.

\section{Diffeomorphisms}

As characterised above, the general covariance of a theory is a matter of the invariance of its equations under smooth but otherwise arbitrary coordinate transformations. In order to make contact with contemporary discussion of the hole argument, we need an alternative formulation that dispenses with reference to coordinates.

Let $g(x)$ stand for some specific solution of a generally covariant theory $T$, expressed with respect to some specific coordinate system $\{x\}$. Let $g'(x')$ be a redescription of the same situation but given with respect to a new coordinate system $\{x'\}$. Since $T$ is generally covariant, $g'(x')$ will also satisfy $T$'s equations (provided that the coordinate transformation $x' = f(x)$ is smooth).

Understood in this way---as a map between descriptions of the very same physical situation given with respect to different coordinate systems---the transformation $g(x) \mapsto g'(x')$ is a \emph{passive transformation}. The \emph{active interpretation} of the transformation involves asking what the function of coordinates $g'$ represents \emph{when interpreted with respect to the original coordinate system}, $\{x\}$. (Note that this question presupposes something that one might dispute, namely, that it makes sense to talk of holding a specific coordinate system fixed, independently of any solution described with respect to it.)

When $g(x) \neq g'(x)$, $g(x)$ and $g'(x)$ will describe \emph{different} physical situations (assuming no redundancy in the way the mathematical object $g$ represents physical reality). And because of $T$'s general covariance, $g'$'s status as a solution of $T$ is independent of which coordinate system, $\{x\}$ or $\{x'\}$, it is referred to. 

The distinction between active and passive transformations first arises in the context of coordinate transformations. The terminology is now used, however, in more general contexts. In particular, one may distinguish between what are labelled (somewhat misleadingly) active and passive \emph{diffeomorphisms}.

Recall that solutions of GR are $n$-tuples, $(M, g_{ab}, \ldots)$, where $M$ is a differentiable manifold. To call $M$ \emph{differentiable} just means that it is equipped with structure that distinguishes, e.g., smooth from non-smooth curves. A \emph{diffeomorphism} between differentiable manifolds $M$ and $N$ is a bijective map such that both the map and its inverse preserve such structure (e.g., under the map, the image of a smooth curve in $M$ will be a smooth curve in $N$ and \emph{vice versa}).

Let $d$ be a diffeomorphism from $M$ to itself. There is no sense in which $d$ \emph{by itself} can be said to be active or passive: it simply associates to each point of $M$ a (possibly distinct) point. To get further, we need to: (i) consider maps naturally associated with $d$ that act on structures defined on $M$ and (ii) distinguish between two types of structure.

The first type of structure includes coordinate systems: maps from $M$ into $\mathbb{R}^4$ (assuming that $M$ is 4-dimensional) that preserve $M$'s differentiable structure. (In fact, the differentiable structure of $M$ is standardly defined extrinsically, via a set of preferred coordinate systems.) The second type includes objects defined on $M$ that are intended to represent something physical: fields or other objects located in spacetime, or physically meaningful spatiotemporal properties and relations.

The definitions of the natural maps associated with $d$ on such objects will differ in detail, depending on the type of object. The basic idea, however, is straightforward. If $d$ maps $p \in M$ to $q$, we can define $d$'s action on an object $F$ via the requirement that the image object, $d^*\!F$, ``takes the same value'' at $p$ as the original object takes at $q$. So, for example, given a coordinate system $\phi: U \subseteq M \to \mathbb{R}^4$, we can define a new coordinate system $d^*\phi$ on the open set $d^{-1}(U)$ via the requirement that $d^*\phi(p) = \phi(d(p))$ for all $p \in d^{-1}(U)$.

A \emph{passive diffeomorphism} corresponds to the case where one contemplates the action of the diffeomorphism on a coordinate system (or systems) whilst leaving the objects representing physically significant structure unchanged. It is the exact correlate of the passive coordinate transformation described above.

An \emph{active diffeomorphism}, by contrast, leaves coordinate systems unchanged but acts on the objects representing physical structure. So, for example, if $(M, g_{ab})$ is a manifold equipped with a metric tensor field $g_{ab}$, $(M, d^*g_{ab})$ is the (in general) mathematically distinct object that results from applying to $g_{ab}$ the active diffeomorphism $d$. $(M, g_{ab})$ and $(M, d^*g_{ab})$ are mathematically distinct because, in general, when $d(p) \neq p$, the value of $d^*g_{ab}$ at $p$ (I write: $d^*g_{ab}|_{p}$) does not equal the value of $g_{ab}$ at $p$.

With this machinery in place, a revised, coordinate-free notion of general covariance, often labelled \emph{(active) diffeomorphism invariance}, can be stated. Let the models of a theory $T$ be $n$-tuples of the form $(M, O_1, O_2, \ldots)$. $T$ is generally covariant if and only if: if $(M, O_1, O_2, \ldots)$ is a structure of the relevant type and $d$ is a diffeomorphism between $M$ and $N$, then $(M, O_1, O_2, \ldots)$ is a solution of $T$ if and only if $(N, d^*O_1, d^*O_2, \ldots)$ is also a solution of $T$. GR is generally covariant in this sense. (Whether this coordinate-free notion of general covariance is equivalent to the notion of general covariance given earlier in terms of coordinate transformations is a subtle business. For more discussion, see \citet{pooley15bac}, where the definition of diffeomorphism invariance is also refined to take account of the distinction between dynamical and non-dynamical fields.)

Finally, I define a \emph{hole diffeomorphism}. Let $M$ be a differential manifold and let $H$ (the ``\emph{hole}'') be a compact open subset of $M$. A diffeomorphism $d: M \to M$ is a hole diffeomorphism corresponding to $H$ if and only if $d$ is the identity transformation outside of $H$ but comes smoothly to differ from the identity transformation within $H$. In other words: $d(p) = p$ for all $p \in M \setminus H$, but $d(p) \neq p$ for some $p \in H$.

\section{The Argument}\label{s:argument}

The core conclusion of the hole argument is that, under a substantivalist interpretation, any generally covariant theory such as GR is indeterministic. (To simplify exposition, I focus on GR as the paradigm generally covariant theory.) There is then the further question of whether this core conclusion counts against a substantivalist interpretation of GR: is the indeterminism in question problematic?

The argument for the core conclusion has three main premises: a claim about what substantivalism entails, a claim about what general covariance entails, and a claim about what it takes for a theory to be deterministic. The core conclusion can be resisted by calling into question each of these claims. They therefore merit careful individual articulation. Before that, however, a brief initial statement of the argument will help ensure that we do not lose sight of the wood for the trees.

Let $\mathcal{M} = (M,g_{ab},T_{ab})$ be a solution of GR. Let $d^*\mathcal{M} = (M,d^*g_{ab},d^*T_{ab})$, where $d$ is a diffeomorphism from $M$ onto itself. It seems that a substantivalist should take $\mathcal{M}$ and $d^*\mathcal{M}$ to represent distinct possibilities. This is because (i) the substantivalist regards the points of $M$ as representing genuine entities, namely substantival spacetime points, and (ii) $\mathcal{M}$ and $d^*\mathcal{M}$ assign the very same points different properties. If $p$ is such that $d(p) \neq p$, then (assuming $g_{ab}$ possesses no symmetries) $\mathcal{M}$ and $d^*\mathcal{M}$ will ascribe different geometrical properties to $p$ ($g_{ab}|_p \neq d^*g_{ab}|_p$) and may ascribe different matter content to (the neighbourhood around) $p$ (if $T_{ab}|_p \neq d^*T_{ab}|_p$).

We assumed that $\mathcal{M}$ is a solution of GR. It follows from GR's general covariance that $d^*\mathcal{M}$ is also a solution. This seems to mean that the (accepting the above reasoning, distinct) situations that $\mathcal{M}$ and $d^*\mathcal{M}$ represent are both physically possible.

Finally, suppose that $d$ is a hole diffeomorphism. In particular, suppose that $M$ is foliable by a family of achronal (with respect to $g_{ab}$) 3-dimensional surfaces, let $\Sigma$ be one such surface, and let $H$ lie entirely to the future of $\Sigma$. (A foliation of an $n$-dimensional manifold is family of disjoint $(n-m)$-dimensional submanifolds ($m < n$) whose union is $M$. In this case, we take $m=1$. A surface is achronal if no two points in the surface lie to the past or future of each other.) $\mathcal{M}$ and $d^*\mathcal{M}$ represent distinct situations ($d$ is non-trivial within $H$) but, outside of $H$, they are identical: they assign exactly the same (spatiotemporal and material) properties to all the points of $M \setminus H$. In particular, the possible spacetimes that they represent are identical up to the instant corresponding to $\Sigma$ but differ to its future. Since both spacetimes are physically possible according to GR, it seems that GR is indeterministic: fixing the laws and the entire spacetime up to $\Sigma$ fails to fix what will happen (which points will have which properties) to the future of $\Sigma$.

That concludes our initial statement of the hole argument. Note that the indeterminism is (in a certain sense) radical. $H$ can be freely specified. Similarly, $d$ can be freely specified, so long as it preserves $M$'s differentiable structure and so long as it smoothly reduces to the identity outside of $H$. This means that, for any compact region of $M$ as small as one likes, completely specifying the properties of spacetime and matter outside of that region fails to fix the properties within the region \citep[524]{earmannorton87}.

The rest of this section develops a slightly more careful version of the argument, designed to be immune to some of the less telling criticisms found in the literature.

I stated above that the core of the hole argument involved three main premises. In the presentation just given, these are: (1) that the substantivalist is committed to taking $\mathcal{M}$ and $d^*\mathcal{M}$ to represent distinct situations; (2) that GR's general covariance entails that the possibilities represented by $\mathcal{M}$ and $d^*\mathcal{M}$ are equally physically possible; and (3) that a theory is indeterministic if it both regards $\mathcal{M}$ and $d^*\mathcal{M}$ as representing distinct possibilities (in the way described) and regards the possibilities as equally possible. For each premise, we should identify the most defensible version that is strong enough to play the required role in the argument.

Premise~(1) is closely related to what Earman and Norton dubbed the ``acid test'' of substantivalism. They wrote:
\begin{quote}
  If everything in the world were reflected East to West (or better, translated 3 feet East), retaining all the relations between bodies, would we have a different world? The substantivalist must answer yes since all the bodies of the world are now in different spatial locations, even though the relations between them are unchanged. (521)
\end{quote}
They then went on to claim that the diffeomorphism ``is the counterpart of Leibniz' replacement of all bodies in space in such a way that their relative relations are preserved'' and concluded that substantivalists were necessarily committed to the denial of \emph{Leibniz Equivalence}, which they defined as the thesis that \emph{diffeomorphic models represent the same physical situation} (522). Here two models $(M, O_1, O_2, \ldots)$ and $(N, O'_1, O'_2, \ldots)$ are diffeomorphic just in case there is a diffeomorphism ${d: M \to N}$ such that, for each object $O_i$, $O_i' = d^*O_i$.

Although Earman and Norton conclude by making a claim about how substantivalists must interpret diffeomorphic models, the Leibniz-inspired scenario that they use to introduce their acid test \emph{makes no mention of models}. Instead, their claim is directly about the physical situations that such models represent. They assert that, for any given situation, a substantivalist must recognise as genuinely distinct the situation where the entire material content of the universe is shifted three feet East relative to its location in the first situation. This very natural assumption went unquestioned by \emph{both} the antisubstantivalist Leibniz and the substantivalist Clarke, in their famous \emph{Correspondence} \citep{leibnizclarke17a-c}. If spatial locations are autonomous entities in their own right, doesn't one have to allow that two situations might be genuinely distinct in virtue of differing only in terms of which substantival places serve as the locations of various material bodies, even if everything else about the two situations is identical?

Premise~(1) is, therefore, best thought of as the combination of two theses: one about the plurality of possibilities that it is alleged a substantivalist must acknowledge; and another about how particular mathematical objects represent those possibilities. Ultimately, it is only the first thesis that does essential work in the hole argument.

Let's label the two theses \textbf{Plurality} and \textbf{Models} and state them more carefully.

\begin{description}
\item[Plurality] Suppose that $P$ is a possible spacetime. The substantivalist is committed to a plurality of possibilities distinct from $P$ that (i) involve the same pattern of spatiotemporal properties instantiated in $P$ and contain the same material fields as $P$, but that (ii) differ from $P$ (solely) over which spacetime points have which properties and serve as the locations of the common material content.
\end{description}

Now for \textbf{Models}. Suppose that $\mathcal{M} = (M,g_{ab},T_{ab})$ can be taken to represent a possible spacetime $P$, and suppose that $P'$ is a distinct but related possibility of the type contemplated in \textbf{Plurality}. Further, suppose that, while differing over how the common geometrical and material properties are distributed over their common set of spacetime points, $P$ and $P'$ do not differ over which collections of points count as smooth paths (i.e., they agree on differentiable structure). The second thesis in Premise~(1) is that, for some suitable choice of diffeomorphism $d$, $d^*\mathcal{M} = (M,d^*g_{ab},d^*T_{ab})$ must be interpreted as representing $P'$.

It is immediately clear, however, that this claim is needlessly strong. It is sufficient for the hole argument that $d^*\mathcal{M}$ \emph{may} be so used. In other words, it is sufficient that there is a permissible \emph{joint} interpretation of the models $\mathcal{M}$ and $d^*\mathcal{M}$ according to which $\mathcal{M}$ represents $P$ and $d^*\mathcal{M}$ represents $P'$. The advocate of the hole argument can easily concede that $\mathcal{M}$ and $d^*\mathcal{M}$ are equally apt to represent either possibility, i.e., that they have the same ``representational capacities'' \citep[332]{weatherall16reg}. No more is required in order to articulate the argument than the following claim:
\begin{description}
\item[Models] \emph{If} $\mathcal{M} = (M,g_{ab},T_{ab})$ can be chosen to represent a possible spacetime $P$ then, \emph{relative to that choice}, there is a permissible and natural interpretation of $d^*\mathcal{M}$ according to which it represents a distinct possibility $P'$.
\end{description}

Let us now turn to Premise~(3), before revisiting Premise~(2). In the spirit of the emendation to Premise~(1), note that whether a theory is deterministic is, in the first instance, a matter of the range of situations that it judges to be possible and only secondarily a matter of how models might represent those possibilities \citep[\emph{cf}][118]{brighouse94}. 

Consider two possible spacetimes, $P$ and $P'$, differing in the way just contemplated. That is, suppose that $P$ and $P'$ involve the same global pattern of spatiotemporal properties and the same global pattern of material fields but that the two spacetimes differ, for some of their common spacetime points, over which of those points instantiate which of the properties common to both spacetimes. Further, suppose that $P$ and $P'$ are in every respect identical up to some global spacelike hypersurface and that their region of disagreement is confined to a ``hole'' to the future of that hypersurface.

If $P$ and $P'$ are both physical possibilities according to the theory, then the theory is, in one obvious and natural sense, indeterministic.
A theory will fail to be deterministic if it is consistent with worlds that involve identical pasts but different futures. In the case at hand, fixing the entire past up to some instant (a region where $P$ and $P'$ match perfectly) fails to fix the future: according to the theory, spacetime's continuation might be that of $P$, or it might be that of $P'$. 

Finally, consider Premise~(2) again. GR's general covariance entails that $\mathcal{M}$ is a solution if and only if $d^*\mathcal{M}$ is a solution. What follows concerning the physical possibility (according to the interpreted theory) of the spacetimes that $\mathcal{M}$ and $d^*\mathcal{M}$ may be taken to represent?

Since we are not naively assuming that $\mathcal{M}$ and $d^*\mathcal{M}$ represent unique possibilities, we should not simply assert that both the spacetime represented by $\mathcal{M}$ and the spacetime represented by $d^*\mathcal{M}$ are equally possible according to GR. These definite descriptions do not pick out unique situations. Rather, the natural claim, in light of our reworked Premises~(1) and (3), is the following:
\begin{description}
\item[Copossible] Suppose that $\mathcal{M}_1$ and $\mathcal{M}_2$ are both solutions to a theory $T$. If there is a permissible joint interpretation of $\mathcal{M}_1$ and $\mathcal{M}_2$ according to which $\mathcal{M}_1$ represents possibility $P_1$ and $\mathcal{M}_2$ represents possibility $P_2$ then if $P_1$ is physically possible according to $T$ so is $P_2$.
\end{description}

\section{Responses to the Argument}

Our reworked premises entail the hole argument's core conclusion: according to the substantivalist, GR is indeterministic. Responses to the argument divide into those that accept this core conclusion and those that reject it. Responses rejecting the core conclusion can then be classified according to which key premise they reject.

For those who accept the core conclusion, the options are to \emph{reject substantivalism} or to bite the bullet and \emph{accept that GR is indeterministic}. In their paper, Earman and Norton favoured the first position. Determinism, they concluded ``may fail, but if it fails it should fail for a reason of physics, not because of a commitment to substantival properties which can be eradicated without affecting the empirical consequences of the theory'' \citep[525]{earmannorton87}.

According to the traditional, bipartite classification, the alternative to substantivalism is \emph{relationalism}. Relationalists deny the (autonomous) reality of spacetime points and analyse facts about spacetime itself as grounded in facts about spatiotemporal properties and relations instantiated by matter. Relationalism evades the hole argument by lacking the plurality of possibilities allegedly plaguing substantivalism: if spacetime points simply do not exist in their own right, there can be no differences between possibilities that concern only which points have which properties.

If relationalists deny that the manifolds in models of GR have a physical correlate, they owe us a positive alternative picture of what such models should be taken to represent. The simplest option is to view physical fields, not as patterns of properties and relations instantiated by the points of substantival spacetime, but as extended objects in their own right, possessing infinitely many degrees of freedom. The role of the manifold is then to represent the continuity and differentiable structure of the fields themselves, and to encode which pointlike parts of one field are coincident with those of another. There are both philosophers and physicists who count as relationalists in this sense and who, to a greater or lesser degree, endorse the hole argument (see, e.g., \citealp[156]{brown05}; \citealp[1309--10]{rovelli07}).

Some remain as sceptical of relationalism as of substantivalism and have sought a ``third way'' between the two. This was the programme that Earman himself tentatively backed \citeyearpar[208]{earman89} but a genuinely novel reconception of the metaphysics of spacetime remains elusive. 
In particular, various attempts to articulate a ``structural realist'' approach to spacetime arguably collapse into variants of either relationalism or (more frequently) substantivalism \cite[see][]{greavesstruc}.

The other option available to someone who accepts the core conclusion is to bite the bullet. Should substantivalists be embarrassed at being forced to view GR as indeterministic? In one sense the indeterminism is pernicious in that, for every possible spacetime, no matter how small a region one considers, the laws and the rest of spacetime fail to fix the state of that region. In another sense, however, the indeterminism is anodyne. Any two possibilities represented by models that differ by a hole diffeomorphism instantiate the very same global pattern of properties and relations. They are therefore \emph{qualitatively} perfectly alike. Their differences involve only which particular individual spacetime points instantiate which properties. In the terminology of modal metaphysics, the differences between the possibilities are purely \emph{haecceitistic} (\citealp{kaplan75}; see also [Caulton, this volume]).

The substantivalist can urge us to recognise that determinism is not an ``all-or-nothing affair'' \citep[13]{earman86}. Given the past of a spacetime, GR might not fix which future individuals get to instantiate this or that qualitative feature, but it might nonetheless fix which qualitative features get to be instantiated. The substantivalist can claim, therefore, that, for all the hole argument has shown, GR is qualitatively or \emph{physically} deterministic: given the past and the laws, all future physical facts might be fixed. The merely haecceitistic facts that fail to be pinned down do not, this substantivalist argues, count as the kind of features of the world that one should expect physics to have anything to say about \citep{brighouse97}. (Note that the hole argument's failing to show that GR is physically indeterministic does not entail that GR is in fact physically deterministic. See \citet[\S 6]{earman07asp} for a review of the wider question of whether GR is deterministic in senses other than that at stake in the hole argument.)

A further step would be to \emph{reject} the notion of determinism presupposed in the hole argument. One would then block the core conclusion of the argument by denying Premise~(3). \citet{leeds95}, for example, argues that whether a theory is deterministic is not a matter of which situations the interpreted theory classifies as possible. Instead, he claims, it is a matter of what sentences are provable within the language of the theory. In order for this strategy to work, it would need to be shown that the notion of determinism presupposed in the hole argument is not merely a possibly disfavoured option amongst several but that it is somehow illegitimate. That seems like a tall order. Leeds himself concedes that his proposal can be read as offering just one more definition of determinism and, moreover, one that matches other definitions framed in model-theoretic or possibility-based terms \citep[435]{leeds95}. If the link between substantivalism and indeterminism is to be severed, Premises~(1) and (2) are more promising targets.

Maudlin seeks to evade the core conclusion on the basis of a position he dubs \emph{metrical essentialism} \citep{maudlin89,maudlin90}. Suppose model $\mathcal{M} = (M, g_{ab}, T_{ab})$ is apt to represent a possible spacetime and consider model $d^*\mathcal{M} = (M, d^*g_{ab}, d^*T_{ab})$ for some diffeomorphism $d$. Recall that Premise~(1) of the hole argument was split above into two components: \textbf{Plurality} and \textbf{Models}. Maudlin accepts \textbf{Models} in at least the following sense: he accepts (in fact insists: see \citealp[84]{maudlin89}) that there is a permissible joint interpretation of $\mathcal{M}$ and $d^*\mathcal{M}$ according to which they represent (if one assumes substantivalism) different \emph{ways for the world to be}. (I here borrow terminology from \citet{salmon89the}.) But, according to Maudlin, these ways for the world to be are not both \emph{ways that the world might have been}; they do not both correspond to genuinely \emph{possible} worlds.

Let us stipulate that model $\mathcal{M}$ represents a possible world. The defining commitment of metrical essentialism is that spacetime points bear their geometrical properties and relations essentially. The value of the curvature scalar at the spacetime point represented---or ``named''---by $p \in M$ is, therefore, one of that point's essential properties. Now suppose that $d(p) \neq p$. In that case, $d^*\mathcal{M}$ represents the very same point as having different geometrical properties, for the value of the curvature scalar at $p$ in $d^*\mathcal{M}$ will (in the generic case) be different from its value in $\mathcal{M}$. It follows that, according to the metrical essentialist, $d^*\mathcal{M}$ represents a state of affairs that is not even metaphysically possible.

Note that the initial choice of $\mathcal{M}$ to represent the genuine possibility is arbitrary---consistently with their representational equivalence, one might equally well have chosen $d^*\mathcal{M}$. (This answers \citeauthor{norton89}'s \citeyearpar[63]{norton89} charge that the metrical essentialist has to explain what distinguishes the ``real'' model from ``imposters''.) What the metrical essentialist insists on is that, \emph{relative to that choice} and \emph{relative to a natural and permissible joint interpretation of the models}, $d^*\mathcal{M}$ represents something metaphysical impossible.

\textbf{Models} had to be tweaked so as to be acceptable to metrical essentialists. Something similar is true of \textbf{Plurality}. In one sense, metrical essentialists block the hole argument by rejecting \textbf{Plurality}. Setting aside cases with nontrivial isometries that are not also symmetries of the matter distribution, the metrical essentialist recognises at most one possible world corresponding to a given pattern of metrical and material properties and relations for any given collection of spacetime points.

To characterise the metrical essentialist as rejecting \textbf{Plurality} is, however, in some ways misleading. Maudlin endorses the intuition behind the ``acid test''; he agrees with Earman and Norton that the substantivalist must view a Leibniz-inspired shift of all matter three feet East as generating a genuinely distinct possibility. His dispute with Earman and Norton is over their classification of diffeomorphisms as the natural generalisations of such shifts. Maudlin stresses that Leibniz shifts apply only to the matter of the universe; they leave the geometric properties of the individual spacetime points unaltered. He therefore sees the models $\mathcal{M} = ( M, g_{ab}, T_{ab} )$ and $\mathcal{M}' = ( M, g_{ab}, d^*T_{ab} )$ as representing the proper generalization of Leibniz shifts when the points of $M$ are interpreted as naming the same spacetime points in each model \citep[552--3]{maudlin90}. And, of course, if $\mathcal{M}$ is a solution of GR, then $\mathcal{M}'$ will, in general, not be (when $T_{ab} \neq \mathbf{0}$). Since at most one of the possibilities represented is physically possible (by the lights of GR), their distinctness does not threaten indeterminism.

Although $\mathcal{M}'$ does not represent a physically possible spacetime, Maudlin will judge that it does represent a \emph{metaphysically} possible spacetime. The atypical case of spacetimes with symmetries is therefore revealing. In such cases, if $d$ is an isometry, $\mathcal{M}'$ can represent a physically possible world genuinely distinct from that represented by $\mathcal{M}$ but one nevertheless qualitatively indiscernible from it. Maudlin thus accepts the meaningfulness of merely haecceitistic distinctions even if he denies that they (invariably) entail a plurality of genuine possibilities.

This suggests the following representation of the metrical essentialist's position. They accept:
\begin{description}
\item[Plurality$^{\mathbf{*}}$] Suppose that $P$ is a possible spacetime. The substantivalist is committed to a plurality of \emph{ways for the world to be} distinct from $P$ that (i) involve the same pattern of spatiotemporal properties instantiated in $P$ and contain the same material fields as $P$, but that (ii) differ from $P$ (solely) over which spacetime points have which properties and serve as the locations of the common material content.
\end{description}
But they reject:
\begin{description}
\item[Copossible$^*$] Suppose that $\mathcal{M}_1$ and $\mathcal{M}_2$ are both solutions to a theory $T$ and let $P_1$ and $P_2$ be ways for the world to be. If there is a permissible joint interpretation of $\mathcal{M}_1$ and $\mathcal{M}_2$ according to which $\mathcal{M}_1$ represents $P_1$ and $\mathcal{M}_2$ represents $P_2$ then if $P_1$ is physically possible according to $T$ so is $P_2$.
\end{description}
With these tweaks, the metrical essentialist counts as someone who accepts Premise~(1) but rejects Premise~(2).

This regimentation highlights that metrical essentialists evade the hole argument's core conclusion only by rejecting what might seem like the obvious moral of the diffeomorphism invariance of the theory. Grant Maudlin that the points of the manifold $M$ can be treated like proper names and that, so understood, models $\mathcal{M}$ and $d^*\mathcal{M}$ represent distinct ways for the world to be. What remains to be decided is whether these ways are both genuinely possible ways for the world to be. Having gone this far, however, it is a very natural further step to take the diffeomorphism invariance of GR as telling us precisely that both states \emph{are} possible. 

In postulating the real existence of spacetime points in the first place, the metrical essentialist is likely to be a scientific realist who is happy to take GR as a guide to ontology. Should not GR also be our guide as to which properties are essential to spacetime points? What the diffeomorphism invariance of GR appears to tell a haecceitist substantivalist is that the only properties essential to a spacetime point are those that it exemplifies as part of a set with the structure of a differentiable manifold \citep[\emph{cf}\/][201]{earman89}.

A different criticism of metrical essentialism focusses on non-isomorphic models. One straightforward way of illustrating the dynamical nature of spacetime structure in GR is to assert, for example, that, had extra mass been present close to some point, then the curvature at that point would have been different \citep[201]{earman89}. How are metrical essentialists to evaluate such counterfactuals? 
They have to judge that the idea that the curvature could have been different from its actual value at \emph{this very point} is as metaphysically absurd as, for example, the idea that Tim Maudlin might have been made of iron girders. Maudlin is forced to concede that ``no model not isometric to the actual world can represent how \emph{this} space-time might have been'' \citep[89--90]{maudlin89} but he insists that dynamically allowed models of GR that are not isometric to the actual world can represent genuine possibilities: ``they are just different possible space-times, not different possible states of this space-time'' \citeyearpar[90]{maudlin89}. He goes on to allow that such possible spacetimes can be used to give a counterpart-theoretic explanation of the truth of Earman's counterfactual. As Brighouse observes \citeyearpar[119--20]{brighouse94}, this is a rather unsatisfying blend of essentialism and counterpart theory.

In the face of such criticism, what positive reasons can metric essentialists offer for their position, aside from the \emph{ad hoc} benefit that it avoids the indeterminism of the hole argument? Maudlin's answer appeals to Newton. In a somewhat obscure passage, which has inspired almost as many different interpretations as commentators, Newton wrote:
\begin{quote}
The parts of duration and space are understood to be the same as they really are only because of their mutual order and position; nor do they have any principle of individuation apart from that order and position, which consequently cannot be altered. \citep[25]{newton04dg}
\end{quote}
According to Maudlin's gloss, Newton is saying that ``the parts of space and time, being intrinsically identical to one another, [have] to be differentiated by their mutual relations of position. Parts of space bear their metrical relations essentially''
\citep[86]{maudlin89}.

This is not especially compelling. Why should someone otherwise comfortable with haecceitistic distinctions think that intrinsically identical objects require (metaphysical?)\ ``differentiation''? Rather than cleaving to haecceitism and avoiding indeterminism by way of an otherwise unmotivated essentialism, perhaps the substantivalist does better to embrace wholeheartedly the ``structuralist'' view that others  \citep[e.g.,][272]{stein02} read in Newton's cryptic remarks. If spacetime points are only ``individuated'' one from another by their spatiotemporal relations (i.e., by their positions in the overall network of spatiotemporal relations), a possible spacetime is exhaustively specified by a complete catalogue of the qualitative facts concerning the full pattern of spatiotemporal relations that are instantiated by its points. According to this antihaecceitist (or \emph{generalist}) point of view, there simply are no further ``individualistic'' facts concerning which objects possess which properties. At a fundamental level, reality, at least concerning spacetime points, is purely qualitative.

Despite the important differences between them, \citet{butterfield89}, \citet{maidens92}, \citet{stachel93,stachel02,stachel06a}, \citet{brighouse94}, \citet{rynasiewicz94}, \citet{hoefer96}, \cite{saunders-stachel}, \citet{pooley06} and \citet{esfeldlam08} all endorse some kind of antihaecceitism, at least concerning spacetime points, whether on general philosophical grounds (as in Hoefer's case), or as a perceived lesson of the diffeomorphism invariance of the physics (as in Stachel's case). Whether acknowledged or not, these authors, in their commitment to spacetime points as entities not reducible to matter and its properties, count as substantivalists, albeit of a ``sophisticated'' variety \citep[228]{belotearman01}.

Sophisticated substantivalists reject the core conclusion of the hole argument by rejecting Premise~(1). In particular, they reject \textbf{Plurality}. The distinct possibilities countenanced by \textbf{Plurality} are precisely possibilities that differ merely haecceitistically. That antihaecceitism and \textbf{Plurality} are incompatible is therefore immediate. One strand of criticism questions whether it is coherent to combine acceptance of spacetime points as entities in their own right with a denial that there are the substantive facts (about which such entities possess which properties) that would generate haecceitistic distinctions. Some argue that a fleshed-out metaphysical story explaining how this combination is possible is still to be given \citep[130--5]{dasgupta11}.

In addition to \textbf{Plurality}, sophisticated substantivalists also reject \textbf{Models}, but this is a simple consequence of their rejection of \textbf{Plurality}, which \textbf{Models} presupposes. This might lead one to wonder whether there is a satisfactory response to the hole argument that rejects \textbf{Models} while disavowing metaphysics and remaining neutral with respect to \textbf{Plurality}. One can interpret \citet{weatherall16reg} and \citet{fletcher18on-} as defending positions of this kind.

According to \textbf{Models}, $\mathcal{M}$ and $d^*\mathcal{M}$ can be used to jointly represent physically distinct situations. The essence of both Weatherall's and Fletcher's views is that this use is not consistent with treating them as Lorentzian manifolds. Weatherall's starting point is that the physical interpretation of a theory's formalism should be consistent with our best understanding of the mathematics of that formalism. In particular, the models employed in a physical theory should count as (physically) equivalent just when they are equivalent according to the mathematics used in formulating those models \citep[331]{weatherall16reg}. Since isometry provides the standard of `sameness' in the mathematics of Lorentzian manifolds, it is a condition on any acceptable interpretation that it regard isometric manifolds, such as $\mathcal{M} = (M,g_{ab})$ and $d^*\mathcal{M} = (M,d^*g_{ab})$, as physically equivalent.

In order to bite against \textbf{Models}, this stricture needs to be understood, not merely as insisting that any two isometric models are equally apt to represent any given possibility (something our formulation of the hole argument was careful to allow), but as ruling out a \emph{joint} interpretation of them on which they are physically inequivalent in the sense that (so interpreted) they represent distinct physical possibilities.

Fletcher is explicit that such a use of the models is in conflict with treating them as members of the mathematical category of Lorentzian manifolds. Any aspect of a state of affairs that is represented by one such model \emph{so conceived} must, he argues, be similarly represented by each isomorphic model. This is because isomorphic models are equivalent ``as objects in that category'': their being isomorphic just is a matter of there being a bijective map of a specific sort that preserves all of the structures constitutive of objects of that type. The consequence Fletcher draws is that any putative representational differences between such isomorphic models are ``not reflected at all in the models themselves as members of [the] category they are taken to be [members of]---there is no mathematical correlate of those differences definable in the category'' \citep[239--40]{fletcher18on-}.

For the sake of argument, let us concede to Weatherall and Fletcher that, on a natural understanding of GR's mathematical machinery, it cannot be used to represent haecceitistic differences between possible spacetimes. What follows for the hole argument? It becomes evident that all the heavy lifting in Premise~(1) is done by its metaphysical component, namely, \textbf{Plurality}. That thesis was not the outcome of a naive way of thinking about the mathematics of GR. It arose from an interrogation of the substantivalist's metaphysics. Weatherall's and Fletcher's reflections, therefore, leave it untouched. 

In effect, both authors argue that, if there are pluralities of merely haecceitistically distinct possibilities, the mathematical formalism of GR, correctly interpreted, is necessarily indifferent to differences between them. But this just means that GR does not distinguish between any two elements of such a plurality; both will count as physically possible according to GR or neither will. And that, of course, is just to admit that, according to any metaphysical view committed to such pluralities, GR is indeterministic. The indeterminism cannot be avoided by remaining loftily above the metaphysical affray.

\section{Defining Determinism}

Contrast the following two definitions of determinism for a theory $T$:
\begin{description}
\item[Det1] $T$ is deterministic just in case, for any worlds $W$ and $W'$ that are possible according to $T$, if the past of $W$ up to some timeslice in $W$ is intrinsically identical to the past of $W'$ up to some timeslice in $W'$, then $W$ and $W'$ are intrinsically identical.

\item[Det2] $T$ is deterministic just in case, for any worlds $W$ and $W'$ that are possible according to $T$, if the past of $W$ up to some timeslice in $W$ is qualitatively (intrinsically) identical to the past of $W'$ up to some timeslice in $W'$, then $W$ and $W'$ are qualitatively identical.
\end{description}
(On these definitions, determinism is a matter of whether the entire history of a world up to some time fixes its future, given the laws. Alternative notions of determinism are easily obtained by considering whether a world at a time (or some other part of a world) fixes the remainder, given the laws. There are also reasons to focus on the conditions for a world, rather than a theory, to be deterministic \citep[468]{brighouse97}. For reasons of space, I ignore these complications.)

According to sophisticated substantivalists, there are no primitive trans-world facts about which objects in one world are identical to which objects in another. On their view, two possible worlds (or two proper parts of distinct worlds) that are not (intrinsically) identical differ qualitatively. Sophisticated substantivalists therefore interpret \textbf{Det1} and \textbf{Det2} as strictly equivalent.

According to straightforward (haecceitist) substantivalists, in contrast, worlds $W$ and $W'$ can differ not just by failing to be perfectly qualitatively alike but by failing to have the very same individuals playing identical qualitative roles. For them, therefore, \textbf{Det1} describes a criterion for determinism that is strictly stronger than that described in \textbf{Det2}. Absent sufficiently strong essentialist constraints on what is possible for spacetime points, straightforward substantivalists will judge GR to be indeterministic according to \textbf{Det1}.

\textbf{Det2} corresponds closely to the definition of determinism offered by David Lewis \citeyearpar[359--60]{lewis83}. A related model-theoretic definition was defended by \citet{butterfield89a}, who argued that it captures the notion of determinism implicit in physicists' discussions of GR's determinism. A significant strand of the hole argument literature has targeted definitions akin to \textbf{Det2}, arguing that they misclassify as deterministic theories that are clearly indeterministic. Such criticism is an obvious problem for sophisticated substantivalists, for whom \textbf{Det2} is equivalent to \textbf{Det1}, but it is also a problem for straightforward substantivalists who, as noted above, might wish to distinguish a notion of physical determinism from determinism \emph{tout court}. \textbf{Det2} might have seemed to adequately capture the former notion.

Problem cases for \textbf{Det2} were raised by \citet[216]{wilson93the} and \citet[418]{rynasiewicz94}, and have been discussed in detail by \citet{belot95}, \citet{brighouse97} and \citet{melia99}. Here are two simple illustrative examples. In the first (adapted from \citealt[660--1]{melia99}) our theory, $T_1$, governs the behaviour of two types of particle: A particles and B particles. Consider a world that contains one A particle equidistant from two B particles, all at rest with respect to one another. Suppose that $T_1$ determines that, at some fixed and predictable time, the A particle will move at a fixed velocity towards one of the B particles. Intuitively, $T_1$ is indeterministic because, despite fixing the qualitative evolution of the situation just described, it fails to fix which B particle the A particle will move towards.

Imagine that we can label the particles in our toy world: $a_1$, $b_1$ and $b_2$. There appear to be two possible futures: one where $a_1$ moves towards $b_1$ and a second where $a_1$ moves towards $b_2$. Haecceitists will judge that $T_1$ is indeterministic according to \textbf{Det1}, for they recognise the merely haecceitistic distinctions between its being $b_1$ or $b_2$ towards which the A particle moves. According to \textbf{Det2}, however, the world is compatible with $T_1$'s being deterministic: if there are two possible futures, they are qualitatively identical.

In our second example (\emph{cf} \citealt[191--2]{belot95}, and \citealt[646--7]{melia99}), theory $T_2$ governs the decay of A particles into B particles. We suppose that everything qualitative about such decays (their spacetime locations, the momenta of the decay products, etc.)\ is fixed by the qualitative history of the world prior to the decay. Now consider a world governed by $T_2$ involving the simultaneous decay of two A particles each into a B particle. The haecceitist will judge that $T_2$ is indeterministic because, despite fixing the qualitative behaviour of all decays, it fails to fix the identities of the decay products. Again, imagine that we can label the four particles and suppose that, in the world we are considering, $a_1$ decays into $b_1$ and $a_2$ decays into $b_2$. A world where $a_1$ decays into $b_2$ and $a_2$ decays into $b_1$, but where everything else is otherwise held fixed, might seem to be an alternative possibility compatible with $T_2$. While $T_2$ is therefore deterministic according to \textbf{Det2} (the qualitative nature of all decays is fully determined by the qualitative nature of the pre-decay state), a haecceitist will judge that the theory fails to be deterministic according to \textbf{Det1}.

Some philosophers \citep[e.g.][]{belot95} accept that both $T_1$ and $T_2$ manifest genuine indeterminism. They have reason to reject \textbf{Det2} and can rest content with a haecceitist understanding of \textbf{Det1}. They are likely to accept the hole argument's conclusion that substantivalist GR is indeterministic.

A sophisticated substantivalist, on the other hand, will view the alleged indeterminism of $T_2$ as suspect: the purportedly distinct possibilities involved in the example are merely haecceitistically distinct. For many, however, the intuition that theory $T_1$ is indeterministic is harder to dispel. Is there a principled way for sophisticated substantivalists to acknowledge that $T_1$ is indeterministic but to deny that $T_2$ is indeterministic?

Despite doubts recently expressed by \citet[\S4]{brighouse18con}, it would seem that this can be done. Consider again the three-particle world governed by $T_1$. According to the haecceitist, there are in fact two such possible worlds: one in which $a_1$ moves towards $b_1$ and another in which $a_1$ moves towards $b_2$. According to the antihaecceitist, there is only one such world: it contains an A particle that moves towards one but not the other of two previously qualitatively identical B particles. But the antihaecceitist can (and, indeed, must) recognise two possible futures for two qualitatively identical \emph{proper parts} of this world. Prior to the A particle's starting to move, there were two qualitatively identical but distinct (and overlapping) pairs composed of an A particle and a B particle. For convenience we can imagine labelling them ``\( (a_1, b_1) \)'' and ``\( (a_1, b_2) \)'' but, note, their distinctness involves no haecceitistic presuppositions. It is secured by the distinctness of \(b_1\) and \(b_2\), two particles coexisting in the same world and situated some distance apart from one another. 

$T_1$'s indeterminism can, therefore, be understood in terms of its failure to fix the (qualitative) future of every part of each world that it governs \citep[652]{melia99}. Take our two pairs of an A particle and a B particle. An exhaustive qualitative specification of such a pair up to the time at which the A particle moves will involve the complete specification of the qualitative history of the whole world up to that time together with a qualitative characterisation of the pair's situation in this history. Up until the time at which the A particle moves, both pairs will satisfy exactly the same qualitiative description. Such a specification, therefore, fails to determine whether one of the particles in the pair will move towards the other or not.

Note that indeterminism is still conceived, as it must be for the antihaecceitist, as a matter of the qualitative past failing to fix the qualitative future. In order to recognise $T_1$'s indeterminism, one just needs to attend to proper parts of a world, in addition to the world as a whole. It turns out that it is relatively straightforward to provide alternative definitions of determinism that regiment the intuitions just described (see \citealt[191, Definition 2]{belot95}; \citealt[\S4.1]{melia99}). This need not mean that \textbf{Det2} should simply be jettisoned. Following 
\citet{dewar16sym}, one might go on to distinguish ``determinism \emph{de dicto}'' (captured by \textbf{Det2}) from ``determinism \emph{de~re}'':
\begin{quote}
\ldots with a little hindsight, it is \emph{utterly unsurprising} that there should turn out to be two concepts of determinism. Determinism is a matter of whether there is one possibility or more consistent with things being a certain way at a certain time; we have two species of possibility, \emph{de dicto} and \emph{de re}; so as a consequence, there are two species of determinism. \citep[53--4]{dewar16sym}
\end{quote}

\section*{Acknowledgements}

I am grateful to numerous colleagues for discussions of the hole argument over the years but, for especially pertinent recent discussion, I am grateful to Eleanor Knox, Sam Fletcher and Jim Weatherall, and, for very helpful comments on an earlier draft, to James Read.

\end{document}